\documentclass[conference]{IEEEtran}

\usepackage{comment}
\usepackage{url}
\usepackage{mathtools}
\usepackage{amsthm}
\usepackage{amsmath}
\usepackage{amssymb}
\usepackage{xfrac}
\usepackage{fontawesome}
\usepackage{bm}
\usepackage{pgfplots}
\usetikzlibrary{calc,external,arrows,arrows.meta,shapes,chains,patterns, fadings,positioning,shadows,decorations.pathreplacing,calligraphy,decorations.text}

\usepackage{algorithm}
\usepackage{algpseudocode}

\usepackage{subfigure}

\usepackage{multirow} 
\usepackage[table,xcdraw]{xcolor}
\usepackage{graphicx}
\usepackage{booktabs}

\graphicspath{{figures/}}
\newcommand{\risk}{\rho}

\begin{document}

\title{Mission-critical spectrum sharing with decentralized Multi-Agent Reinforcement Learning}

\author{
  \IEEEauthorblockN{
    Dimitrios Pylorof\IEEEauthorrefmark{1},
    Imtiaz Nasim\IEEEauthorrefmark{1},
    Humberto E. Garcia\IEEEauthorrefmark{1},
    Vivek Agarwal\IEEEauthorrefmark{1},
    J. A. Mannil\IEEEauthorrefmark{2}, and
    Mingyue Ji\IEEEauthorrefmark{3}
  }
  \\
  \IEEEauthorblockA{
    \IEEEauthorrefmark{1}\textit{Idaho National Laboratory}, USA, \{Dimitrios.ParsinasPylorof, Imtiaz.Nasim, Humberto.Garcia, Vivek.Agarwal\}@inl.gov\\ 
    \IEEEauthorrefmark{2}\textit{Howard University}, \{jasni.mannil \}@bison.howard.edu\\
    \IEEEauthorrefmark{3}\textit{University of Florida}, \{mingyueji\}@ufl.edu\\
  }
}

\maketitle

\begin{abstract}
Motivated by emerging mission-critical applications and an increasingly congested spectrum, we develop a decentralized multi-agent reinforcement learning (MARL) model for dynamic spectrum access. The model enables secondary users to learn effective transmission strategies across shared frequency bands while minimizing collisions with high-priority primary users and among themselves. We design the agent-level learners following a Markov potential game approach, connecting independent local updates to system-level improvement. We instantiate this design using lightweight linear actor--critic learners suitable for resource-constrained edge devices, rather than computationally intensive centralized or deep multi-agent architectures. Across spectrum environments with different incumbent activities, the learned policies adapt their transmission policy and waiting behavior to preserve throughput while greatly reducing transmission collisions relative to random and forecast-aware heuristic baselines. The results establish the value of decentralized MARL and shows up to 96.8\% reduction in overall collisions.

\end{abstract}
\begin{IEEEkeywords}
Spectrum sharing, MARL, Spectrum occupancy
\end{IEEEkeywords}

\IEEEpeerreviewmaketitle

\section{Introduction}
Spectrum scarcity is increasingly a problem of coordination rather than a simple absence of nominal bandwidth. Primary users (PUs) involving federal, public-safety, energy, transportation, and defense systems may have priority access, while secondary systems seek intermittent opportunities without disrupting incumbent missions. \textit{Cognitive radio} and \textit{dynamic-spectrum-access (DSA)} research has therefore emphasized sensing, prediction, band selection, and distributed access control \cite{akyildiz2006next}. 
However, DSA involves more than identifying an empty band only as incumbents pose bursty activities over time. Secondary users (SUs) may select the same band using forecasting, and it can still be inaccurate at the exact instant of transmission \cite{lopez2013time,mosavat2021prediction}.

Reinforcement learning (RL) offers a natural way for learning spectrum-access decisions and their long-term consequences.  Prior works have explored deep multiband access and distributed multi-user policies,  \cite{wang2018deep,naparstek2019deep} as well as DQN/DRQN spectrum sharing and actor--critic designs that jointly learn band access and transmission-power decisions \cite{liu2023dynamic,dong2022dynamic}. More recent studies have focused on practical decentralized access. Fair Share RL learns fairness-oriented band selection policies from local transmission outcomes without inter-agent coordination \cite{zhang2025fair}. Similarly, a sparse recurrent multi-agent prunes deep networks within an independent actor/global-critic architecture to reduce deployment cost \cite{stamatelis2025multiagent}. These results reinforce the value of learned distributed access, but leave an important research gap for distinct mission-critical spectrum access: lightweight decentralized actor-critic learning must operate under imperfect long-horizon forecasts while providing operator-readable methods for incumbent protection. In particular, a key operational behavior is that an intelligent node must learn not only  over which band to transmit, but also \emph{when not to transmit}. Deferring certain transmissions can not only protect a primary incumbent, but also reduce contention among peer SUs.

This paper develops and evaluates that ``value of waiting'' in a concrete decentralized MARL implementation. While the broader framework provisions for active sensing and adaptive transmit-power control as natural extensions, this work deliberately restricts the learning agents to two actions:- remaining idle or fixed-power band selection . The scope of this paper investigates a foundational question first: \emph{can decentralized MARL produce meaningful coordination and collision reduction before the action space is enlarged?} We provide reasonable details with validated results to answer this question in the following sections.

Our main contributions are:
\begin{itemize}
\setlength{\itemsep}{0pt}
\setlength{\parskip}{0pt}
  \item We formulate forecast-aware decentralized access with a finite action policy per agent: idle or transmit at a fixed power on one of the available bands.
  This implementation isolates the coordination learned through waiting and fixed-power band selection.

  \item We evaluate our MARL method against several strong benchmarks and demonstrate its effectiveness under varying incumbent activity.
  \item Our results identify conservative operating points that maintain non-trivial service, while decision diagnostics quantify when the agents should wait and when to reject high-risk immediate transmissions. Results show that the proposed MARL model reduces PU-SU collisions up to $96.8\%$ compared to forecast-aware heuristic baseline.
\end{itemize}
Spectrum awareness alone does \textit{not} provide spectrum coordination. When several secondary users identify the same apparently favorable band, forecast-aware access can simply synchronize their contention. The proposed decentralized MARL instead learns the temporal and spectral coordination needed to decide not only where to transmit, but whether each \textit{current} opportunity is worth taking.

\section{Background and Design Position}
\subsection{Distributed learning and potential-game structure}
A finite Markov game contains agents $i\in\{1,\ldots,N\}$, local observations $o_i$, actions $a_i$, joint action $\bm a$, transition kernel, and individual rewards $r_i$. Independent learners treat the policies of other agents as part of a changing environment \cite{zhang2021marl}, which is useful for decentralized deployment.

Potential games provide a useful structure for aligning individual agent decisions with system-level objectives to satisfy
\begin{equation}
 r_i(s,a'_i,a_{-i})-r_i(s,a_i,a_{-i})=\Phi(s,a'_i,a_{-i})-\Phi(s,a_i,a_{-i}),
 \label{eq:potential}
\end{equation} 
for a common potential $\Phi$ \cite{monderer1996potential}. Markov potential game (MPG) extends this alignment to long-term rewards, showing that independent policy updates can improve a common system objective under suitable assumptions \cite{leonardos2022global,ding2022independent}. Recent constrained-MPG results make the safety connection more explicit. Alatur et al. formulate coupled policy constraints and learn Nash policies in tabular finite-horizon games \cite{alatur2024provably}, while Jordan et al. develop independent policy-gradient learning for approximate constrained Nash equilibria \cite{jordan2024independent}. PU protection can naturally be formulated as a cumulative interference or collision constraint. Our independent policy gradient MARL implementation in Section \ref{sec:DPG} combines this perspective with lightweight independent policies and linear function approximation. We follow MPG design principles through finite actions, independent local actors and critics, common reward formulations, and queue transitions driven by local actions and exogenous arrivals. 

\subsection{Forecasting as a look-ahead sensor}
Incumbent prediction can support proactive access decisions, but continuously forecasting recieved signal strength (RSS) is harder than a binary occupancy label. Previous studies such as \cite{wu2021autoformer} use an Autoformer to separate the trend and seasonal components for long-horizon forecasting, while \cite{pan2023autoformer} applies an Autoformer-CSA assisted cross-band modeling for spectrum sequences. However, recent work with GAD-Autoformer replaces fixed moving-average decomposition with a gated, stateful trend mechanism that adapts for bursty incumbent activities while retaining longer memory during quiet periods \cite{hu2026gad}. Results in this study \cite{hu2026gad} also revealed that the GAD-Autoformer provides high accuracy to forecast incumbent RSS across different horizons, with the largest advantage observed at long horizons.

Consequently, this paper 
uses the forecaster from \cite{hu2026gad} as a heuristic baseline, and utilizes this as an external sensor for the proposed model. Future RSS values are generated by perturbing an incumbent reference trace according to the empirical residual distribution at each forecast horizon.  Residuals are drawn independently for each agent, resulting in different local forecasts while retaining the same underlying reference trace. This setup enables us to evaluate how effectively distributed agents learn temporal access strategies from realistic but imperfect spectrum forecasts.

\section{Spectrum Sharing Environment}
The environment models SUs sharing incumbent-prioritized bands and sending queued packets to a common sink. SU-to-sink distance determines path loss and received signal power; when multiple SUs use the same band, the resulting signal-to-interference-plus-noise ratios (SINRs) determine successful transmissions or peer collision. At each decision step, every SU receives a noisy look-ahead spectrum forecast and chooses whether to wait or transmit. Incumbent activity governs PU conflicts, while concurrent SU choices govern peer contention.

\subsection{Incumbent and RSS process}
Each band follows a semi-Markov ON/OFF process with Weibull-distributed dwell times, producing persistent, irregular bursts rather than memoryless switching. We consider a  scalar \emph{activity knob} $\kappa$ as in \cite{hu2026gad} that shifts the process from relatively quiet to progressively busier regimes, while time-of-day, weekend, and band-specific variations keep the bands from evolving identically.

The activity knob is a generator control, not a prescribed occupancy fraction. Realized occupancy emerges from the stochastic dwell times and temporal modulation, so $\kappa$ is used throughout the evaluation as a qualitative stress indicator: \emph{quiet} (smaller $\kappa$) $\rightarrow$ \emph{busier} (larger $\kappa$). The latent ON/OFF process is mapped to continuous RSS levels, yielding gradual and noisy changes rather than binary labels.

\subsection{Forecast-error surrogation}
Spectrum awareness enters through an imperfect look-ahead forecast. The simulator models GAD-Autoformer forecast error by perturbing the hidden future RSS with an empirical residual drawn at the corresponding lead. For agent $i$, band $b$, and lead $h$,
\begin{equation}
\widehat P_{i,b}(t+h)=P_b(t+h)+\varepsilon_{i,b,h},
\label{eq:surrogate}
\end{equation}
where $P_b(t+h)$ is the hidden environment RSS and $\varepsilon_{i,b,h}$ is the sampled forecast residual. The policy receives the perturbed forecasts and derived uncertainty features. Agent-specific residuals give different nodes different local views of the same evolving spectrum.

\begin{figure}[h]
\centering
\begin{tikzpicture}
\begin{axis}[
  width=0.90\columnwidth,
  height=3.9cm,
  xmin=-23,xmax=23,
  ymin=0,ymax=0.14,
  xtick={-20,-10,0,10,20},
  ytick={0,0.05,0.10},
  xlabel={Forecast error, estimate $-$ reference (dB)},
  ylabel={Density},
  title={Representative forecast residuals},
  grid=major,
  tick label style={font=\scriptsize},
  label style={font=\scriptsize},
  title style={font=\scriptsize},
]
\addplot[
  thick,
  mark=*,
  mark size=1.15pt,
] coordinates {
(-22,0.001)(-20,0.003)(-18,0.006)(-16,0.007)(-14,0.011)
(-12,0.026)(-10,0.024)(-8,0.018)(-6,0.016)(-4,0.018)
(-2,0.019)(0,0.045)(2,0.080)(4,0.130)(6,0.060)
(8,0.033)(10,0.021)(12,0.011)(14,0.008)(16,0.009)
(18,0.008)(20,0.004)(22,0.003)
};
\addplot[densely dashed,thin] coordinates {(0,0)(0,0.14)};
\end{axis}
\end{tikzpicture}
\caption{Representative GAD-Autoformer \cite{hu2026gad}\textit{ forecast residual density} used to calibrate the spectrum awareness surrogate used in our spectrum sharing environment for MARL training and evaluation.}
\label{fig:forecast-residual}
\end{figure}
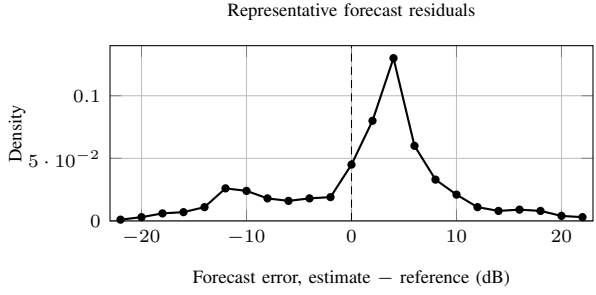

Forecasted RSS is converted to a smooth busy-risk representation and combined with forecast uncertainty before entering the policy features. Across the look-ahead interval, the lead-indexed residual profiles are empirically similar: persistent quiet/busy regimes and recurring burst-transition errors repeatedly produce the same broad error structure. Fig.~\ref{fig:forecast-residual} shows one such representative profile.

\subsection{Actions, queues, and wireless outcomes}
Each SU selects either an idle action or a fixed-power transmission on one available band $b$,
\begin{equation}
\mathcal A_i=\{\mathrm{idle}\}\cup\{\mathrm{TX}(b): b\in\mathcal B\},
\end{equation}
where $\mathcal B=\{1,\ldots,B\}$ denotes the available bands. Traffic queues receive stochastic packet arrivals. Each transmission attempt removes one queued packet; the realized outcome enters separately through the reward. An attempt on an incumbent-active band is a PU conflict. Otherwise, same-band SUs are resolved through the path-loss/SINR model: the strongest link may lead to a successful transmission, while weaker same-band links incur peer collisions. The learning therefore balances incumbent protection against coordination with peers competing for the same favorable bands.

\section{Decentralized Policy Gradient Formulation} \label{sec:DPG}
The learning loop can be read in three stages: each node, which corresponds to a SU, first compresses its queue, time, and imperfect spectrum forecast into a fixed feature vector; it then assigns preferences to its local actions; finally, it adjusts those preferences from its own observed rewards. This setup keeps the relation between forecast information, mission priorities, and learned behavior visible while remaining close to the finite-action, linear-approximation MPG described in Sec.~II. We retain $N$ for the number of agents and $B$ for the number of bands, with index sets $\mathcal N$ and $\mathcal B$. Let $\Delta t$ denote the decision interval, $P_{\rm TX}$ the fixed transmit power, $q_0$ the initial queue length, $q_{\max}$ the queue capacity, and $p_{\rm arr}$ the per-step packet-arrival probability.

\begin{figure*}[h]
\centering
{\footnotesize{\input{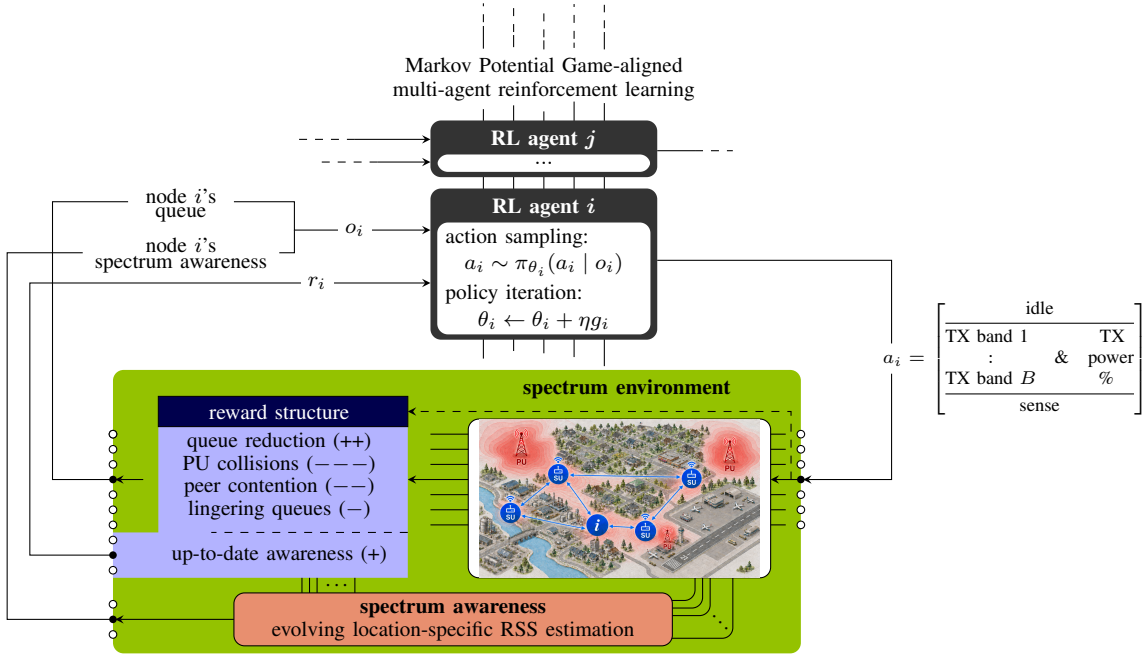}}}
\caption{Conceptual MPG-based spectrum sharing architecture. Adaptive power and sensing actions are marked as additions in follow-on work; this study uses only idle and fixed-20-dBm band selection to first establish the value of decentralized MARL.}
\label{fig:architecture}
\end{figure*}

\subsection{From forecasts to a value-of-waiting signal}
Agent $i$ observes its queue, per-band forecasted RSS and busy probability with uncertainty over a look-ahead horizon of $H$ steps. These form its \textit{local observation} $o_i$. Next, agent $i$ forms its\textit{ local feature vector} $\phi(o_i)$, by parsing $o_i$ as we illustrate next. Our aim is to assemble a feature vector that leads to numerically stable operations downstream in the independent policy gradient algorithm, \textit{and} exposes each agent to the information we wish (or posit) to leverage for attaining intelligent decentralized reasoning. The forecasted RSS is converted to a smooth \textit{busy risk} score using a center $\tau_{\rm RSS}$ and scale $s_{\rm RSS}$ (both of which are selected to match typically observed conditions in the corresponding spectrum environment), via the logistic mapping:
\begin{equation}
\widehat p_{i,b,h}
=\left[1+\exp\!\left(-\frac{\widehat P_{i,b}(t+h)-\tau_{\rm RSS}}{s_{\rm RSS}}\right)\right]^{-1}.
\label{eq:rssrisk}
\end{equation}
The \textit{busy risk} score from (\ref{eq:rssrisk}) is then conservatively adjusted for known or assumed forecast uncertainty, by adding a weighted uncertainty term. Let $\widehat\sigma_{i,b,h}$ denote the associated forecast-uncertainty feature and $\beta$ its risk weight. The risk-adjusted forecast is
\begin{equation}
\risk_{i,b,h}=\widehat p_{i,b,h}+\beta\widehat\sigma_{i,b,h}.
\label{eq:riskadjusted}
\end{equation}
The policy feature map $\phi(o_i)\in\mathbb R^{d_\phi}$ summarizes current per-band risk, the best and average opportunities later in the forecast, which bands look best now and later, and interactions with the current queue. A particularly interpretable feature is
\begin{equation}
\Delta_i^{\mathrm{wait}}(t)=\min_b \risk_{i,b,0}-\min_{b,\,1\le h\le H}\risk_{i,b,h}.
\label{eq:waitadv}
\end{equation}
A positive $\Delta_i^{\mathrm{wait}}$ means that the forecast contains a safer opportunity later than the best band available now. This is only forecast-derived information supplied to the learner: waiting is beneficial only if the expected reduction in interference is worth carrying the queue for longer; the learner still needs to evaluate this tradeoff using our independent policy gradient iterations.

\subsection{Reward weights and mission postures}
Operator preference enters through the relative weighting of reward terms. Successful delivery receives positive value, while PU conflicts receive the largest penalty; peer collisions, link failures, unnecessary attempts, idling, and backlog are weighted separately as shown in the conceptual architecture in Fig. \ref{fig:architecture}. The implemented reward collects these terms as
\begin{align}
r_i(t)={}&\rho_s \mathbb{I}_i^{\mathrm{succ}}-\lambda_{\mathrm{PU}}\mathbb{I}_i^{\mathrm{PU}}-\lambda_{\mathrm{SU}}\mathbb{I}_i^{\mathrm{peer}}-\lambda_{\ell}\mathbb{I}_i^{\mathrm{link}} \nonumber\\
&-\lambda_a \mathbb{I}_i^{\mathrm{att}}-\lambda_0 \mathbb{I}_i^{\mathrm{idle}}-\lambda_q\frac{q_i(t+1)}{q_{\max}}.
\label{eq:reward}
\end{align}
Here, each $\mathbb{I}_i^{(\cdot)}\in\{0,1\}$ is an event indicator for agent $i$. $\mathbb{I}_i^{\mathrm{succ}}$ denotes a successful delivery, $\mathbb{I}_i^{\mathrm{PU}}$ is an attempted transmission on an incumbent-active band, $\mathbb{I}_i^{\mathrm{peer}}$ is a same-band SU collision, and $\mathbb{I}_i^{\mathrm{link}}$ indicates a transmission that fails the link/SINR test without a PU conflict. The action indicators $\mathbb{I}_i^{\mathrm{att}}$ and $\mathbb{I}_i^{\mathrm{idle}}$ mark any transmission attempt and an idle decision, respectively. Thus a transmission attempt incurs the attempt term in addition to exactly one realized transmission outcome. All agents use the same coefficient vector. Changing that vector does not change the learning algorithm; it changes the \textit{mission posture} or  local incentives. We evaluate several alternatives that deliberately emphasize different operating objectives, from strict incumbent protection to more service-seeking behavior. Table~\ref{tab:reward-weightings} summarizes the alternatives MARL postures which are chosen to illustrate how the same decentralized learners adapt their behavior when the mission emphasizes protection, queue relief, or service more strongly.
\begin{table}[t]
\caption{Alternative reward weightings explored to induce different mission postures. All $\lambda$ values are nonnegative penalty weights in \eqref{eq:reward}; $\rho_s$ weights successful delivery.
}
\label{tab:reward-weightings}
\centering
\scriptsize
\setlength{\tabcolsep}{4.0pt}
\begin{tabular}{lccccccccc}
\toprule
Metric & Strict  & Base  & Queue-aware  & Balanced\\
\midrule
$\rho_s$ & 1.2 & 1.5 & 1.5 & 2.3\\
$\lambda_{\rm PU}$ & 10 & 8.0 & 8.0 & 7.0\\
$\lambda_{\rm SU}$ & 0.45 & 0.45 & 0.45 & 0.45\\
$\lambda_{\ell}$ & 0.30 & 0.30 & 0.30 & 0.30\\
$\lambda_a$ & 0.06 & 0.05 & 0.05 & 0.03\\
$\lambda_0$ & 0 & 0 & 0 & 0.005\\
$\lambda_q$ & 0.055 & 0.055 & 0.080 & 0.085\\
PU/success weight ratio ($=\lambda_{\mathrm{PU}}/\rho_s)$ & 8.33 & 5.33 & 5.33 & 3.04\\
Plotted $\kappa$ & .10 & .15, .20 & .05 & .25\\
\bottomrule
\end{tabular}
\end{table}

\subsection{From action scores to independent policy updates}
Once the features and reward are defined, the learner needs to answer two questions: which action should be preferred in the current state, and did the resulting outcome turn out better or worse than expected? We use a small \textit{linear actor} and \textit{critic} and, in adherence with our decentralized reasoning, we pursue both answers locally and directly tied to the particular agent's feature vector. For the actor, each feasible action $a$ receives a linear score $z_{i,a}=\phi(o_i)^{\mathsf{T}}\theta_{i,a}$, where $\theta_{i,a}$ are the actor coefficients for action $a$. A softmax converts the action scores into probabilities, 
\begin{equation}
\pi_{\theta_i}(a\mid o_i)=\frac{\exp(z_{i,a})}{\sum_{a'\in\mathcal A_i}\exp(z_{i,a'})}.
\label{eq:policy}
\end{equation}
Actor coefficients $\theta_{i,a}$ are initialized independently from a zero-mean Gaussian distribution with standard deviation $\sigma_\theta$. 

Learning proceeds from on-policy batches of $M$ transitions. Let $\Phi_i$ and $\Phi'_i$ stack the current and successor feature vectors, $\bm r_i$ stack the rewards, $\gamma$ denote the discount factor, and $\lambda$ the critic regularization coefficient. With $X_i=\Phi_i-\gamma\Phi'_i$, our ridge-regularized least-squares Bellman-residual critic is
\begin{equation}
w_i=(X_i^{\mathsf{T}}X_i+\lambda I)^{-1}X_i^{\mathsf{T}}\bm r_i,
\label{eq:lstd}
\end{equation}
where $I$ is the identity matrix of compatible dimensions. The critic supplies a local expectation of discounted return. Each observed transition is compared with that expectation through the TD residual, limited in magnitude by $d_{\max}$,
\begin{equation}
\delta_{i,t}=\operatorname{clip}\!\left(r_{i,t}+\gamma\phi_{i,t}'{}^{\mathsf{T}}w_i-\phi_{i,t}^{\mathsf{T}}w_i,
[-d_{\max},d_{\max}]\right).
\label{eq:tdres}
\end{equation}
Positive $\delta_{i,t}$ reinforces the sampled action and negative $\delta_{i,t}$ suppresses it. Let $\eta$ denote the actor step size and $g_{\max}$ the gradient-norm limit. The independent policy-gradient estimate and actor update are
\begin{align}
g_i&=\frac{1}{M}\sum_{t=1}^{M}\delta_{i,t}\,
\nabla_{\theta_i}\log \pi_{\theta_i}(a_{i,t}\mid o_{i,t}), \nonumber\\
\theta_i&\leftarrow\theta_i+\eta\,\operatorname{clip}_{g_{\max}}(g_i).
\label{eq:actorupdate}
\end{align}
Thus the critic is used only to judge recent \textit{local} experience; there is no centralized value function or exchange of learned parameters.

Let $T_{\rm train}$ denote the training horizon. Each agent updates after every $M$ newly collected transitions and, after step $T_{\rm train}$, applies the same update once to any residual buffer content before its policy is frozen. Let $T_{\rm eval}$ denote the frozen-policy deployment horizon and $T_{\rm tail}$ the reporting window. For deployment, $\epsilon_{\rm tie}$ denotes the tolerance used to randomize near-tied greedy actions, while $\tau_{\rm eval}$ denotes the softmax temperature for the stochastic alternative. Let $\mathcal S$ denote the evaluation seed set. Algorithm~\ref{alg:training} summarizes this ordering and refers back to Eqs.~\eqref{eq:lstd}--\eqref{eq:actorupdate} for the learning updates.

The learned parameter count follows directly from $d_\phi$ and the action-space size: each node stores $d_\phi|\mathcal A_i|$ actor coefficients and $d_\phi$ critic coefficients. This compact construction keeps the learned action logic inspectable, remains close to the MPG and linear function approximation analysis \cite{ding2022independent}, and is compatible with CPU-class edge execution without a centralized critic.

\begin{algorithm}[t]
\caption{Forecast-aware decentralized MARL training and frozen-policy evaluation}
\label{alg:training}
\footnotesize
\algtext*{EndFor}
\algtext*{EndIf}
\begin{algorithmic}[1]
\Require Settings in Sec.~\ref{sec:settings}; lead-indexed residual libraries $\{\mathcal E_h\}_{h=0}^{H}$
\State Initialize queues, local critics $w_i$, actors $\theta_i$, and empty local buffers $\mathcal D_i$
\For{$t=0,1,\ldots,T_{\rm train}-1$}
  \State Advance the hidden PU/RSS process; draw lead-specific forecast errors and form each agent's local forecast and network-summary proxy
  \ForAll{agents $i$ \textbf{in parallel}}
    \State Construct $\phi_{i,t}$, including risk and $\Delta_i^{\rm wait}$; mask TX if $q_i=0$ and sample $a_{i,t}$ from \eqref{eq:policy}
  \EndFor
  \State Resolve the joint wireless outcome; compute rewards, update queues, form successor features, and append each transition to $\mathcal D_i$
  \If{$|\mathcal D_i|=M$}
    \State \textbf{in parallel for all $i$:} fit the critic by \eqref{eq:lstd}, form TD residuals by \eqref{eq:tdres}, update the actor by \eqref{eq:actorupdate}, and clear $\mathcal D_i$
  \EndIf
\EndFor
\State Update once on any residual buffer content using \eqref{eq:lstd}--\eqref{eq:actorupdate}; freeze the actors
\State Restart the seeded trace, deploy for $T_{\rm eval}$ steps, and report the final $T_{\rm tail}$ steps
\end{algorithmic}
\end{algorithm}

\subsection{$\epsilon$-Nash convergence} \label{sec:RLconvergence} 
Our independent policy gradient construction used in this paper follows the MPG direction of \cite{ding2022independent}, which establishes approximate $\epsilon$-Nash convergence with linear function approximation. 
Under the additional conditions establishing coincidence of Nash equilibria with potential maximizers, the resulting $\epsilon$-Nash convergence can be related directly to approximate maximization of the global spectrum sharing objective.

Comparably with any practical implemented learner, the pursued MPG alignment and associated literature results define the \textit{proof envelope} for our learning architecture.

\subsection{Implementation settings}
\label{sec:settings}
For the numerical instantiation used throughout this paper, $N=6$, $B=4$, $\Delta t=5$ min, and $P_{\rm TX}=20$ dBm, with $\mathcal N=\{1,\ldots,N\}$ and $\mathcal B=\{1,\ldots,B\}$. Hence $|\mathcal A_i|=B+1=5$. Queues start at $q_0=8$ packets, have capacity $q_{\max}=20$, and receive a Bernoulli arrival with $p_{\rm arr}=0.20$ per decision step.

For the forecast and feature map, $H=24$, $\tau_{\rm RSS}=-82$ dBm, $s_{\rm RSS}=4.5$ dB, $\beta=0.75$, and $d_\phi=82$. The forecast spans two hours. Each node stores $410$ actor coefficients and $82$ critic coefficients.

For learning, $\sigma_\theta=0.01$, $M=256$, $\gamma=0.99$, $\lambda=10^{-3}$, $d_{\max}=50$, $g_{\max}=20$, and $\eta=0.015$. Training runs for $T_{\rm train}=3000$ environment steps. This produces twelve updates in total: eleven after full $256$-transition batches and one final update using the $184$ transitions left at the end. For frozen-policy evaluation, $T_{\rm eval}=1000$, $T_{\rm tail}=250$, $\epsilon_{\rm tie}=10^{-3}$, and $\tau_{\rm eval}=0.5$. Reported metrics use the final $T_{\rm tail}$ deployment steps, with event counts scaled to a per-$1000$-step basis.

\begin{figure*}
    \centering
    \includegraphics[width=.98\linewidth]{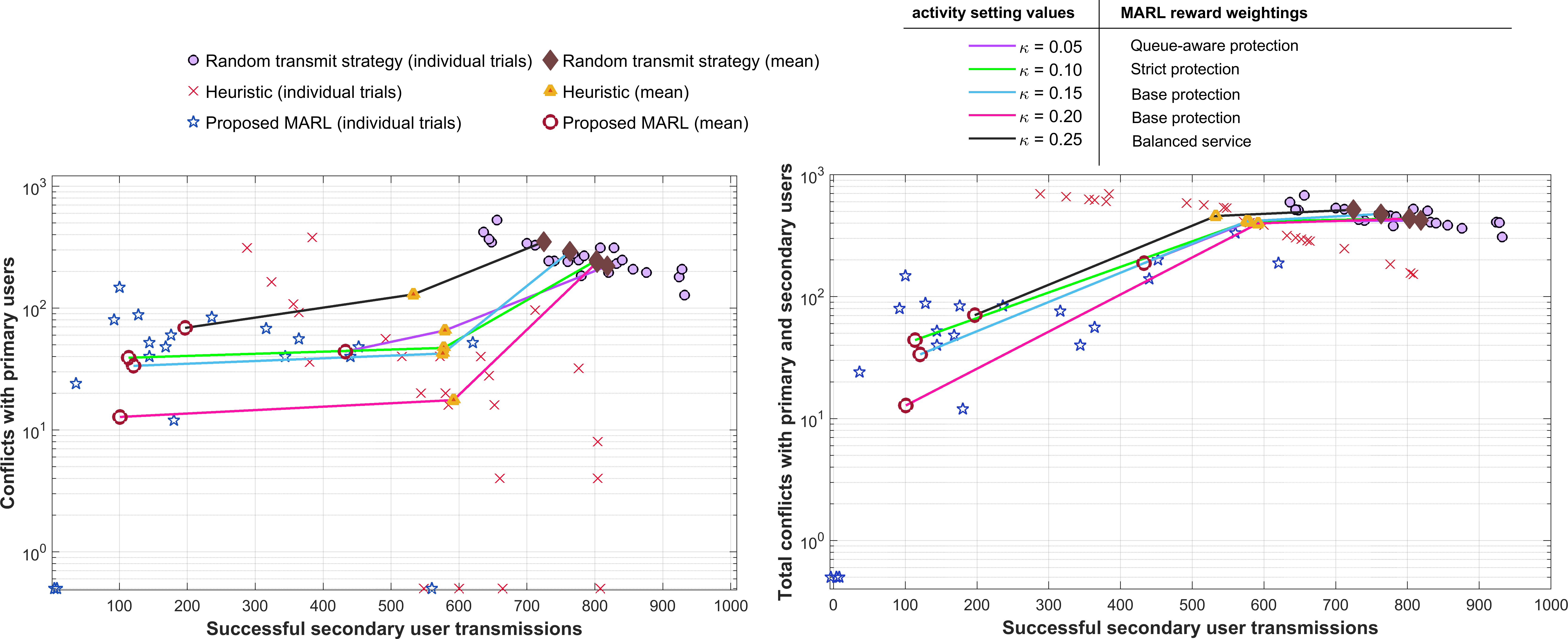}
    \caption{Successful SU transmissions versus PU conflicts (left) and total PU+SU conflicts (right). Faint markers show individual trials, bold markers show five-seed means, and connecting lines identify the activity setting $\kappa$ (indicated and color-coded as \textit{PU occupancy}, for brevity). The value of the selected \textbf{MARL operating points}, and even of the originating distribution of individual trial results --especially in the right plot-- can be seen in laying \textbf{consistently lower} (i.e., indicating lower number of conflicts) than the \textbf{heuristic points} and the \textbf{random baseline points}. The illustrated legends apply to both left and right panels.}
    \label{fig:results_landscape}
\end{figure*}
\section{Numerical Evaluation}

\subsection{Protocol and baselines}
Each activity setting $\kappa \in \{0.05, 0.10, 0.15, 0.20, 0.25\}$ uses a $14$-day trace of our spectrum sharing environment, sampled every five minutes (resulting in $4032$ timestamps). Every command-line sweep trains on-policy for $3000$ environment steps, or $250$ h ($10$ days $10$ h) of simulated operation, using the update staging described in Section~IV-C. After the final training update, the actors are frozen and the environment is restarted for $1000$ deployment steps ($83$ h $20$ min). Methods are evaluated over five different seed values, and reported metrics average the final $250$ steps ($20$ h $50$ min), scaled per $1000$ steps where appropriate. 

We compare the performance of a selected MARL operating point against a random baseline and a forecast-aware heuristic, within an environment defined by each activity setting value. The \emph{random baseline} transmits with probability $0.65$ whenever its queue is nonempty and selects a band uniformly. The \emph{forecast-aware heuristic} computes current risk $\risk_{i,b,0}$, transmits on the minimum-risk band; if this probability is greater  than $0.95$ then stays idle.

After training and evaluation, we use a single diagnostic score to select one representative MARL operating point from the reward-weighting and deployment-mode sweep at each activity setting,
\begin{equation}
J_{\mathrm{score}}=S-3.5C_{\mathrm{PU}}-0.45C_{\mathrm{peer}}-0.30C_{\mathrm{link}}-0.08N_{\mathrm{att}}.
\label{eq:fixedscore}
\end{equation}
Here, $S$ is the number of successful SU transmissions, $C_{\mathrm{PU}}$ the number of PU conflicts, $C_{\mathrm{peer}}$ the number of peer collisions, $C_{\mathrm{link}}$ the number of link/SINR failures, and $N_{\mathrm{att}}$ the total number of SU transmission attempts. All quantities are computed over the same evaluation tail and scaled to a per-$1000$-step basis. This score is used only for post-training selection; it does not enter the agent reward, policy-gradient update, or environment dynamics. Among configurations achieving at least $100$ successful SU transmissions per $1000$ steps, we select the one with the largest $J_{\mathrm{score}}$, breaking ties by throughput. The selected configuration is the MARL point shown in Fig.~\ref{fig:results_landscape} and used in the subsequent comparison tables. In order of increasing activity knob values, the ultimately selected profiles correspond to queue-aware protection, strict protection, base protection, and balanced service from Table~\ref{tab:reward-weightings}.

\subsection{How reward weightings move the learned operating point}
In Fig. \ref{fig:results_landscape} we view the sequence of selected profiles as operationally coherent and only show the MARL posture (See Table~\ref{tab:reward-weightings}) that provides the best performance. At $\kappa=.05$, the MARL postures queue-aware and base profiles are nearly tied; the selected queue-aware point provides a slight upper-hand with $432.8$ successes and $188.0$ total conflicts (right panel in Figure). At $\kappa=.10$, strict protection posture remains service-feasible and reduces the selected point to $44.0$ total conflicts. At $\kappa=.15$, the strict profile produces only $28.0$ successes and fails the service floor, so base protection becomes the best feasible posture. Hence, we plotted the result for base protection only for MARL for $\kappa=0.15$. At $\kappa=.20$, strict protection collapses to complete waiting, while base protection just clears the floor with $100.8$ successes and $12.8$ conflicts. At $\kappa=.25$, strict, base, and queue-aware protection all converge to zero-service deployment; the balanced profile reopens access and yields $196.8$ successes with $70.4$ total conflicts. The final row in Table \ref{tab:reward-weightings} identifies where each mission posture generated the plotted MARL point for Fig. \ref{fig:results_landscape}. Overall, we notice how reward weight selection chooses between recognizable mission postures and reveals a behavioral transition from selective access to stronger deferral when the incumbent-conflict penalty becomes too large.

\begin{table}[t]
\caption{Relative collision reduction of plotted MARL points versus the heuristic (five-seed means per $1000$ steps)}
\label{tab:collision-reduction}
\centering
\scriptsize
\setlength{\tabcolsep}{2.1pt}
\begin{tabular}{c r r r r r r}
\toprule
& \multicolumn{3}{c}{PU conflicts} & \multicolumn{3}{c}{Total PU+SU conflicts}\\
\cmidrule(lr){2-4}\cmidrule(lr){5-7}
$\kappa$ & heuristic & MARL & reduction & heuristic & MARL & reduction\\
\midrule
.05 & 65.6 & 44.0 & 32.9\% & 412.0 & 188.0 & 54.4\%\\
.10 & 47.2 & 39.2 & 16.9\% & 413.6 & 44.0 & 89.4\%\\
.15 & 42.4 & 33.6 & 20.8\% & 415.2 & 33.6 & 91.9\%\\
.20 & 17.6 & 12.8 & 27.3\% & 399.2 & 12.8 & 96.8\%\\
.25 & 129.6 & 68.8 & 46.9\% & 458.4 & 70.4 & 84.6\%\\
\bottomrule
\end{tabular}
\end{table}

\subsection{Throughput--conflict operating regions}
As seen in Fig.~\ref{fig:results_landscape}, random transmit strategy produces high throughput by attempting often, but incurs both PU and peer conflicts. The forecast-aware heuristic strongly reduces PU conflicts, yet all agents tend to identify and select the same attractive band; its peer-conflict burden remains between $328.8$ and $381.6$ per $1000$ steps. The proposed MARL points occupy a different region: throughput is lower, while total conflict burden falls sharply. Table~\ref{tab:collision-reduction} shows PU-conflict reductions of $16.9$--$46.9\%$ and total-conflict reductions of $54.4$--$96.8\%$. At $\kappa=0.20$, MARL meets the considered minimum service with $100.8$ successes, while lowering PU conflicts from $17.6$ to $12.8$. Therefore, the learned MARL configurations populate low-conflict regions that are not reached by random access benchmark and are obscured when the heuristic is judged only by its low PU-conflict count. The total-conflict view is essential because a low-PU heuristic may still coordinate SUs poorly. One key rationale for the outperformance of MARL in Table~\ref{tab:collision-reduction} is--unlike the heuristic solution, the proposed MARL leverages the delayed queue-versus-risk tradeoff induced by the full forecast horizon, and anticipates that several agents may select the same band safely.

\begin{table}[t]
\caption{Access behavior for observations with nonempty queues across the
five activity settings. MARL values show the range across the selected
operating points.}
\label{tab:waiting}
\centering
\scriptsize
\setlength{\tabcolsep}{3.4pt}
\begin{tabular}{lccc}
\toprule
& \multicolumn{2}{c}{Access behavior} & Forecast-responsive behavior \\
\cmidrule(lr){2-3}\cmidrule(l){4-4}
Policy
& Wait
& High-risk TX
& Wait for safer future opportunity \\
\midrule
Random        & 32.5\%       & 57.8\%       & n/a \\
Heuristic     & 62.5\%       & 33.0\%       & n/a \\
Selected MARL & 84.3--98.1\% & 1.9--15.7\%  & 35.6--40.4\% \\
\bottomrule
\end{tabular}
\end{table}

\subsection{What the policy learns about waiting}
Effective conflict reduction requires not only appropriate band selection but also learning when to defer transmission. Table~\ref{tab:waiting} illustrates when waiting reduces risk and can be more attractive than selecting any band for transmission. Selected MARL policies wait in $84.3$--$98.1\%$ of these states, reduce high-risk transmissions to $1.9$--$15.7\%$, and wait for
a forecast improvement in $35.6$--$40.4\%$ of queued states. The main cost of this conservative behavior is increased backlog: mean queue lengths are $10.8$--$18.7$ packets for MARL, compared with $10.7$ for the heuristic and about $0.38$ for random access. Waiting is \textit{not always} preferable: the reward weights determine how strongly the learned policy trades incumbent protection and peer coordination against queue buildup and associated latency, while the minimum service requirement excludes operating points that achieve low conflict rates simply by remaining idle.

\section{Discussion and Roadmap}
\label{sec:roadmap}

Decentralized spectrum access requires more than accurate spectrum awareness. Even when individual nodes identify low-risk bands, independent access decisions can concentrate secondary users onto the same opportunities and create substantial peer contention. The proposed MARL provides a complementary coordination layer: agents learn to reach low-conflict operating regions by deciding when to access, which band to use, and when to defer (i.e., wait). Communication network operators can deliberately shift the resulting operating point across successful transmissions, PU conflicts, and SU conflicts by tuning the MARL reward weights. Meanwhile, the lightweight decentralized learners can be deployed on CPU-class edge devices and learn policies adapted to the spectrum environment experienced locally.

\textbf{Distributed awareness.} Next, each node will update a local GAD-Autoformer from its own, locally-sampled RSS history, producing location-specific embeddings and uncertainties. Unlike the work in this paper, the future scope will consider nodes that can exchange compact summaries over a low-rate control band, leverage a distributed average consensus mechanism, and collaboratively build and maintain joint spectrum awareness.

\textbf{Richer actions and theory boundary.} An expanded action space can include the learnable action to perform sensing during a particular step and contribute to renewed spectrum awareness (in addition to either staying idle or transmitting), as well as transmitting more than one packet at selectable power levels. This will also introduce additional energy use and conservation considerations. 
In parallel, another interesting scope is to replace the soft PU reward weight with an explicit horizon-wise conflict or interference budget, moving the formulation toward constrained MPGs \cite{alatur2024provably,jordan2024independent}. All such directions would still require new analysis beyond current guarantees, perhaps justifiably pushing beyond the MARL/MPG frontier. Our approach in this work, however, aimed to first establish that lightweight forecast-aware MARL is worth building upon, and then explore the following challenges while tracking which MPG assumption is the best fit.

\section{Applications of proposed decentralized spectrum coordination for response operations}
The proposed decentralized spectrum sharing, and, in a broader sense, coordination approach is directly relevant to diverse civilian and consumer infrastructures. In addition, it is of interest to homeland security and emergency response networks, in which heterogeneous mobile, maritime, airborne, and uncrewed assets must communicate under congested, contested, or satellite-denied conditions. Representative applications include Coast Guard underway and over-the-horizon communications, as well as police and Border Patrol tactical networks, where communication resources are distributed, centralized control may be unreliable or infeasible, and protection of higher priority spectrum users is important. These mission settings align well with three properties of our approach: each node makes access decisions locally without a centralized critic, the learned policy is sufficiently compact for CPU-class edge hardware, and forecast-aware waiting provides an explicit mechanism for trading service urgency against incumbent protection and peer contention.

An appealing transition path is to embed the learned access policy as a contention-management layer within nodes of an existing tactical network or traffic manager. The present results establish mechanism viability and promise, rather than fielded performance: operational deployment would require representative RF testing, and subsequent network-level demonstrations with relevant edge platforms. Such steps are critical to determine whether the safety--service operating regions observed in this paper persist under realistic node topologies and geometries, interference, and real-world communication and spectrum challenges.
\section{Conclusion}
We presented forecast-aware decentralized MARL for primary-protective spectrum sharing. Six agents used imperfect long-horizon RSS forecasts to choose waiting or one of four fixed-power bands. Lightweight local actor/critic updates provided low-conflict operating points with a nonzero service floor and no centralized policy or critic. Our key result shows that spectrum availability should not lead to inefficient spectrum access decisions. Forecasting can identify an apparently safe transmission opportunity, but decentralized MARL can learn whether taking that opportunity is beneficial once incumbent risk, peer contention, and future opportunities are considered jointly. The resulting policies occupy low-conflict operating regions that forecast-aware access alone does not meet.

\section*{Acknowledgment}
This work was supported by the Idaho National Laboratory Laboratory Directed Research and Development program under U.S. Department of Energy Idaho Operations Office Contract DE-AC07-05ID14517, and, in part, by the U.S. Department of Homeland Security Science and Technology Directorate and the ADAC-ARCTIC Center of Excellence managed by the University of Alaska Anchorage. 
The views and opinions of authors expressed herein do not necessarily state or reflect those of the U.S. Government or any agency or Contractor.

\bibliographystyle{IEEEtran}
\bibliography{references}

\end{document}